\documentclass[12pt]{iopart}

\newcommand{\ind}{\hspace*{1in}}
\usepackage{iopams}
\usepackage{graphicx}
\usepackage{cite}

\begin{document}

\title[]{Study of van der Waals bonding and interactions
in metal organic framework materials}

\author{Sebastian Zuluaga} \address{Department of Physics, Wake
Forest University, Winston-Salem, NC 27109, USA.}

\author{Pieremanuele Canepa} \address{Department of Physics, Wake
Forest University, Winston-Salem, NC 27109, USA.} 

\author{Kui Tan} \address{Department of Materials Science and
Engineering, University of Texas at Dallas, TX 75080, USA.}

\author{Yves J. Chabal}
\address{Department of Materials Science and
Engineering, University of Texas at Dallas, TX 75080, USA.}

\author{Timo Thonhauser} \ead{thonhauser@wfu.edu}
\address{Department of Physics, Wake Forest University,
Winston-Salem, NC 27109, USA.}


\begin{abstract}
Metal organic framework (MOF) materials have attracted a lot of
attention due to their numerous applications in fields such as hydrogen
storage, carbon capture, and gas sequestration. In all these
applications, van der Waals forces dominate the interaction between the
small guest molecules and the walls of the MOFs. In this review article,
we describe how a combined theoretical and experimental approach can
successfully be used to study those weak interactions and elucidate the
adsorption mechanisms important for various applications. On the theory
side, we show that, while standard density functional theory is not
capable of correctly describing van der Waals interactions, functionals
especially designed to include van der Waals forces exist, yielding
results in remarkable agreement with experiment. From the experimental
point of view, we show examples in which IR adsorption and Raman
spectroscopy are essential to study molecule/MOF interactions.
Importantly, we emphasize throughout this review that a combination of
theory and experiment is crucial to effectively gain further
understanding. In particular, we review such combined studies for the
adsorption mechanism of small molecules in MOFs, the chemical stability
of MOFs under humid conditions, water cluster formation inside MOFs, and
the diffusion of small molecules into MOFs. The understanding of these
phenomena is critical for the rational design of new MOFs with desired
properties.
\end{abstract}

\maketitle

\section{Introduction}

Metal organic framework (MOF) materials are nano-porous materials
comprised of metal centers, which are linked by organic ligands. Over
the past decade, MOFs have attracted a surge of attention due to their
extraordinary properties, useful for hydrogen storage \cite{h_storage,
h_storage2, h_storage3, h_storage4}, CO$_2$ capture \cite{CO2_capt,
CO2_capt2, CO2_capt3, CO2_capt4, Pera_2013}, catalysis
\cite{Farha_MOFcatal2, Lee_MOFcatal3, Shultz_MOFcatal}, and sensing
\cite{Kreno12} among others \cite{Stroppa13, Stroppa11}.  Part of the
success of MOFs also has to do with their often simple synthesis, i.e.\
by combining the organic ligands and the metallic salt in a solvothermal
reaction \cite{mof_app, mof_prep}. Most practical applications of MOFs
rely on a specific interaction of the MOF with small molecules. This
interaction---typically of a weak van der Waals type---has thus been at
the center of many experimental and theoretical studies. It is exactly
the understanding of this interaction that will allow us to interpret
the properties of current MOFs better and design new and improved MOFs
with desired properties. For example, we do know that, in general, the
MOFs surface area and the binding strength to the metal centers are the
two main factors controlling the uptake of small molecules.  However,
the exact correlation between those properties is unclear
\cite{h_storage5}. Another example concerns much current research,
trying to address the problem of low stability of MOFs under humid
conditions. While some progress has been made \cite{Yang_2013,
Taylor_2013, Li_2013, Han_2010, Demessence_2009, wikipaper55}, the newly
found water-resistant MOFs often lack the desired specific molecular
uptakes that are needed. Overall, progress has been slow to address such
questions due to a lack of appropriate methods to study the molecular
interactions inside MOFs. In the present review article we highlight a
strategy, combining experiment and theory, that overcomes these problems
and has been particularly successful in unraveling van der Waals
interactions in MOFs.

The experimental study of those interactions often relies on powerful
vibrational spectroscopy such as infrared (IR) absorption and Raman
scattering, which indirectly provide information about the molecular
adsorption process taking place in the MOF. The theoretical description
with \emph{ab initio} methods, due to the typical size of MOF unit cells
and their extended nature, rules out most highly accurate
quantum-chemistry approaches and leaves density functional theory (DFT)
as the only viable option. Historically, however, standard
exchange-correlation functionals within DFT such as LDA and GGA only
poorly capture van der Waals interactions. We will show here that the
specially designed functional vdW-DF \cite{VdW1, VdW2, VdW3} is in fact
capable to describe van der Waals interactions reliably and gives
results in remarkable agreement with experiment.

This review article aims to showcase the importance of IR and Raman
spectroscopy techniques combined with \emph{ab initio} simulations at
the DFT level (utilizing vdW-DF) as a promising way to study and
rationally design complex systems where van der Waals bonding plays a
major role. To this end, this work is divided into several sections. In
Section \ref{sec:IR_Raman}, we give a description of the success and
failures of vibrational spectroscopic techniques to study van der Waals
interactions. Then, in Section \ref{sec:comp}, we present a description
of the computer simulations used to describe and interpret complex
spectroscopic experiments. In Section \ref{sec:cases}, we present
several relevant examples where the combination of experiment and theory
explains the behavior of various MOF systems and provides much needed
understanding. We conclude with a short summary and outlook in Section
\ref{sec:summary}.

\section{Success and Failure of Vibrational Spectroscopies to study van
der Waals Interactions}\label{sec:IR_Raman}

\subsection{IR and Raman spectroscopy of small molecule adsorption in MOFs}

IR and Raman spectroscopy provide complementary information about
bonding configurations through their vibrational spectra. IR spectra
reflect photon absorption during transitions from ground- to
first-excited vibrational levels ($\nu =0 \to 1$) in the electronic
ground state, requiring a dynamic dipole moment (associated with a
change in the dipole moment during the vibrational
motion) \cite{Nakamoto_2009_UTD1}.  In contrast, Raman spectroscopy is
based on photon scattering by molecules and has its origin in the
electronic polarization caused by monochromatic visible radiation
\cite{Nakamoto_2009_UTD1, Ferraro_2003_UTD2}. Therefore, a vibrational
mode is Raman active if the polarizability is modulated during the
vibration \cite{Nakamoto_2009_UTD1, Ferraro_2003_UTD2}. Strict selection
rules exist for both spectroscopies, sometimes leading to complementary
detection \cite{Ferraro_2003_UTD2}.  For example, the vibration of the
homopolar diatomic molecule H$_2$ is not IR active (due to the absence
of a fluctuating dipole associated with the symmetric stretching), but
strongly Raman active. However, once the molecule interacts with the
MOF, it undergoes a perturbation that slightly polarized the originally
symmetric molecule and makes it weakly IR active. This perturbation is
usually accompanied by a red-shift of the H--H stretching modes, located
at 4161.1~cm$^{-1}$ and 4155~cm$^{-1}$ for para and ortho H$_2$,
respectively \cite{Welsh_1969_UTD3}. For the linear molecule CO$_2$, the
symmetric stretch mode ($\nu _{1}$) is Raman active but not IR active,
whereas the antisymmetric modes ($\nu _{2}$ and $\nu _{3}$) are IR
active \cite{Ferraro_2003_UTD2}.

Based on these principles, IR and Raman spectroscopy can be very useful
tools to characterize the nature of host/guest
interactions \cite{Lamberti_2010_UTD4, Vimont_2007_UTD5,
Gascon_2009_UTD6, Stavitski_2011_UTD7} in MOFs. Particularly valuable
information can be gained by identifying perturbations of the IR active
modes. For example, the first spectroscopic evidence for the formation
of an electron-donor acceptor (EDA) complex between CO$_2$ and
functional groups of MOFs was observed in a MOF of type MIL--53 and
reported in later studies of adsorption of CO$_2$ in amino-based
MOFs \cite{Vimont_2007_UTD5, Gascon_2009_UTD6}. The adsorption of CO$_2$
molecules in MIL--53 leads to a modest red-shift from $-10$~cm$^{-1}$ to
$-15$~cm$^{-1}$ of the stretching mode $\nu _{3}$ and to a splitting of
the bending mode $\nu _{2}$ due to the removal of degeneracy of the
in-plane and out-of-plane bends \cite{Vimont_2007_UTD5}. A similar
splitting of $\nu _{2}$ modes is common in many
electron-donor acceptor complexes of CO$_2$ with organic solvents or
polymers possessing electron-donating functional groups---e.g., carbonyl
groups---due to the interaction of the carbon atom of
CO$_2$ as the electron acceptor \cite{Dobrowolski_1992_UTD8,
Kazarian_1996_UTD9}. Moreover, significant perturbations of both
$\nu$(OH) and $\sigma$(OH) bands of hydroxyl groups ($\nu$(OH) = 19
cm$^{-1}$ and $\sigma$(OH) = 30 cm$^{-1}$) suggest that oxygen atoms of
the framework hydroxyl group act as the electron
donor \cite{Vimont_2007_UTD5}.

As evident from these examples (and many others), it is clear that IR
and Raman spectroscopy, by themselves and even without the aid of
theoretical calculations, can often provide insight into the
interactions between guest molecules and the MOF. However, as we will
see in the next section, in other cases the ``blind'' application of
these spectroscopic techniques can lead to a significant
misinterpretation of the experimental data obtained. This can happen
when IR and Raman spectroscopy are used as \emph{indirect}
probes---i.e.\ deducing other physical properties of the system from a
simple red- or blue-shift in the spectrum. In such cases, theory and
computer simulations are essential to derive a complete understanding,
as they provide \emph{direct} access to many properties of the system,
and often provide interpretations that are unexpected from simple
correlations in the experimental data.

\subsection{Difficulty of IR and Raman spectroscopy to describe small
molecule adsorption in MOFs}

Despite the high sensitivity of spectroscopy to molecular interactions
with the MOF, attention must be paid when interpreting the data to
extract information about the interaction from vibrational frequency
shifts, intensities, and line-widths \cite{Nijem_2010_UTD10,
Nijem_2010_UTD11}. For example, it is commonly accepted that the
magnitude of the IR shift of small adsorbed molecules in MOFs is
directly related to their adsorption energy, and thus the IR shift can
be used indirectly to estimate the relative adsorption energies.
However, in our recent IR spectroscopy study of molecular hydrogen in a
number of different MOF compounds \cite{Nijem_2010_UTD10} , we find that
there is no clear correlation between H$_2$ adsorption energies
(determined by isotherm measurements) and the magnitude of the H$_2$
stretch shift. In fact, metal-formate M$_3$(HCOO)$_6$ [M = Co, Ni and
Mn] compounds with the highest adsorption energy have the lowest
hydrogen IR shift. In this case, we find that the IR shift is dominated
by the environment (organic ligand, metal center, and structure)
\cite{Nijem_2010_UTD10}, rather than by the adsorption energy to the
metal.

Similarly, integrated areas for the specific IR bands were long
considered to be directly correlated with the amount (loading) of
adsorbed molecules, assuming that the dipole moment of the adsorbed
species is not affected by loading or site geometry
\cite{Bordiga_2007_UTD12, Vitillo_2008_UTD13, Garrone_2005_UTD14}. Based
on this assumption, variable temperature IR was used to measure the
absorbance of IR bands (including that of H$_2$ molecules) and estimate
the adsorption energy \cite{Bordiga_2007_UTD12, Garrone_2005_UTD14}.
However, our theoretical and experimental findings for H$_2$ molecules
in MOF74 with unsaturated metal centers indicate that large variations
in the induced dipole moment take place as a function of loading, due to
the interaction among adsorbed molecules \cite{Nijem_2010_UTD11}. In the
case of Mg-MOF74, the effective charge of H$_2$ at the metal sites
weakens (from 0.021~$e$ to 0.015~$e$ as the loading increases from 1 to
12 H$_2$/primitive cell) i.e. as the neighboring sites are occupied.
Thus, the IR intensity is reduced by 50$\%$, since it is proportional to
the square of the effective charge or the dynamic dipole
moment \cite{Nijem_2010_UTD11}.  These findings suggest that the
integrated areas of IR bands do not always correlate with the amount of
H$_2$ adsorbed and possible variations in dynamic dipole moments have to
be taken into account. 

In summary, IR and Raman spectroscopy can be very helpful tools when
studying small molecule adsorption in MOFs. But, extreme caution is
necessary when utilizing those methods to make assumptions about
adsorption energies or loadings, as illustrated in the examples given
above. Under these circumstances, theoretical input using first
principles calculations---specifically capable to deal with van der
Waals interactions---is critical to interpret experimental observations
correctly.

\subsection{Experimentation}

Zecchina and coworkers first used transmission IR spectroscopy to study
the fundamental aspects of the interaction between H$_2$ and MOFs,
mainly in the low temperature ($<$300~K) and pressure regime. By means
of the variable temperature infrared (VTIR) spectroscopy method, the
adsorption enthalpy was derived by measuring the intensity of absorption
bands as a function of temperature \cite{Bordiga_2005,
Vitillo_2008_UTD13}. However, caution must be taken when using the VTIR
method since the dipole moment might change as a function of loading, as
pointed out above. More recent work \cite{Nijem_2010_UTD10,
Nijem_2010_UTD11, wikipaper55, wikipaper45} has investigated a series of
small molecules (H$_2$, CO$_2$, CH$_4$, SO$_2$, H$_2$O, etc.) using {\it
in situ} IR absorption spectroscopic to quantify the effect of their
interaction with different types of MOFs in a wide range of pressures
(from 50 mTorr to 55 bar) and temperatures (10 K to 423 K). In order to
perform the IR measurements at and above room temperature, a portion
($\sim$10 mg) of MOF was lightly pressed onto a KBr support and mounted
into a high-temperature high-pressure cell (Specac product number P/N
5850c) and further heated in vacuum for activation. During the
annealing, the removal of solvent molecules was monitored by {\it
in situ} IR spectroscopy. Then, the activated sample was cooled to
specific temperatures in order to perform the measurements at specific
pressure gas exposures. Measurements were performed in transmission
using a \mbox {liquid--N$_2$} cooled InSb/MCT detector. Similar
measurements were also performed in a Janis PTSHI series cold
refrigerator (CCR) system for low temperature studies ($<$298~K). In
addition to transmission IR, diffuse reflectance infrared Fourier
transform spectroscopy (DRIFTS) was employed to investigate the dynamics
of H$_2$ molecules adsorbed within the MOF74 compounds
\cite{Gascon_2009_UTD6, Stavitski_2011_UTD7}.  Furthermore, DRIFTS has
also been used to study the interactions between CO$_2$ and functional
groups on the organic ligands of some MOFs under the controlled {\it in
situ} cell environment \cite{FitzGerald_2011, FitzGerald_2010, Windisch_2009}. 

Most recently, {\it in situ} Raman spectroscopy was also used to study
the structural response mechanism of flexible metal organic frameworks
Zn$_2$(bpdc)$_2$bpee [bpdc = 4,40-biphenyl dicarbox-ylate and bpee =
1,2-bis(4-pyridyl)ethylene] upon CO$_2$, N$_2$, and hydrocarbon
molecules adsorption \cite{CO2_capt8, wikipaper47}. In this case, Raman
spectroscopy is more suitable because the phonon modes of the MOFs do
not overwhelm the spectra as they do in the case of IR spectra, which
include a large number of combination and overtone bands. By integrating
a Linkam FTIR600 cooling/heating stage, the activated sample was
measured under the controlled temperature and gas environment. The
changes on specific bonds in the MOF structure, monitored by Raman
spectroscopy, were correlated to the MOF structural changes and the
guest-host interactions.

\section{Computer Simulations as a Tool to Interpret Complex
Spectroscopic Experiments}\label{sec:comp}

\subsection{Ab initio modeling of materials}

While very successful classical modeling techniques exist, such as force
field simulations, which are suitable for studying very large systems,
they are not capable of describing the electronic structure of materials
and the intricate role it plays in many processes. In the case of MOF
materials, we are most interested in electronic-structure changes during
the adsorption and desorption of small molecules in their cavities, as
well as a number of catalytic processes. As such, unless cost
prohibitive, \emph{ab initio} modeling techniques are the methods of
choice. For an overview of widely-used materials-modeling techniques,
ranging from classical approaches to high-level quantum-chemistry
methods, see Ref.~\cite{Kolb2012a}.

Modeling MOF materials with \emph{ab initio} methods presents a
particular challenge. The adsorption/desorption of small molecules in
MOFs is often governed by physisorption, i.e.\ weak van der Waals
forces, which are difficult to capture correctly with \emph{ab initio}
methods. Correlated high-level quantum-chemistry approaches, such as
M{\o}ller-Plesset perturbation theory and coupled-cluster methods
\cite{Szabo_96}, can describe van der Waals interactions, but their
computational cost limits them to small systems ($\sim$100 and $\sim$30
atoms, respectively \cite{Head-Gordon}) and their application to large
periodic systems, such as the MOFs of interest here, is unpractical
\cite{Marsman_2009, Booth_2012, Gruneis_2010, Usvya_2011, Ayala_2001,
Maschio_2007}.

Density functional theory (DFT) \cite{Kohn_64}, on the other hand, has a
much more favorable computational cost and can be used for systems with
up to 1000 atoms---in linear-scaling DFT even up to 1,000,000 atoms
\cite{Head-Gordon}. It is also easily implemented with periodic
plane-wave basis sets, such that treating periodic systems becomes
trivial. Unfortunately, with standard exchange-correlation functionals,
DFT cannot reliably describe van der Waals
interactions~\cite{Shevlin_09, Hobza_08, Sponer_08, Cerny_07}, a
phenomenon where charge fluctuations in one part of the system correlate
with fluctuations in another, resulting in an attractive force that is a
\emph{truly nonlocal correlation effect} \cite{Langreth_09}. It follows
that standard local and semi-local functionals, such as LDA and GGA,
cannot reliably account for these nonlocal effects and yield
qualitatively erroneous predictions \cite{French_10, Kristyan_94,
Perez-Jorda_94, Meijer_96}. While very promising extensions exist
\cite{French_10}, most notably DFT-D~\cite{Grimme1, Grimme2},
DFT-SAPT~\cite{Hesselmann_03, Misquitta_02, Williams_01}, and
$C_6$-based methods \cite{Tkatchenko_09, Tkatchenko_09_2}, they are
semi-empirical, perturbative, and not seamlessly self-consistent density
functionals.

\subsection{vdW-DF: a good compromise between cost and accuracy}

We have overcome this problem and include van der Waals forces
self-consistently in DFT \cite{VdW2} in the form of a van der Waals
density functional (vdW-DF). Its accuracy is comparable to high-level
quantum-chemistry approaches \cite{Cooper_08, Copper_08_2, Li_08}.
vdW-DF goes beyond standard DFT to include a \emph{truly nonlocal
correlation} $E^{\rm nl}_c$ in the exchange-correlation energy $E_{xc}$,
\begin{eqnarray}
\label{equ:functional}
E_{xc}[n] &=& E^{\rm revPBE}_x[n]  + E^{\rm LDA}_c[n] +
   E^{\rm nl}_c[n]\;,\\[2ex]
E_c^{\rm nl}[n] &=& \frac{1}{2}\int d^3r\,d^3r'\,\,
   n(\vec{r}\,)\phi(\vec{r},\vec{r}\,')n(\vec{r}\,')\;,
\end{eqnarray}
where $n$ is the electron density and revPBE~\cite{revPBE} and
LDA~\cite{LDA} are standard functionals. $E_c^{\rm nl}$ is determined by
the kernel $\phi$, which is a complicated function of the densities and
their gradients at $\vec{r}$ and $\vec{r}\,'$, first developed by
Langreth et al.\ \cite{VdW1}. With the corresponding
exchange-correlation potential $v^{\rm nl}_c(\vec{r}\,) = \delta E^{\rm
nl}_c[n]/ \delta n(\vec{r}\,)$ \cite{VdW2}, the method becomes
self-consistent and permits calculation of atomic forces---essential for
structural optimization and molecular-dynamics simulations. Like
high-level quantum-chemical methods, vdW-DF precludes bias through a
full, self-consistent solution of the coupled Schr\"odinger equations
for all valence electrons. Calculations automatically include direct and
induced electrostatic effects, charge transfer effects, and effects due
to the non-nuclear centricity of the dispersion interaction as well as
its deviations from the inverse sixth power at smaller than asymptotic
separations.

vdW-DF can be implemented in standard plane-wave electronic-structure
codes exploiting an efficient Fast Fourier Transform algorithm
\cite{Soler_09}. This algorithm scales like standard DFT, and for large
systems compute times are negligibly longer than for GGA calculations.
We routinely use our own implementation to study hydrogen-storage
materials with 300 to 400 atoms per unit cell \cite{Bil_11,
our_wiki_40}. In particular, we have successfully used vdW-DF to study a
variety of phenomena in MOF materials, achieving remarkable agreement
with experiment \cite{wikipaper43, wikipaper45, wikipaper47,
wikipaper48, wikipaper50, Piero_PRL, wikipaper51, wikipaper53,
wikipaper55, Tan_2013, Canepa2013}.

\subsection{Comparison between vdW-DF simulations and experiment}

In MOF adsorption studies, there are ample opportunities for theory and
experiment to interact. It is almost straight forward to compare vdW-DF
optimized structures with diffraction experiments \cite{Walker_2010,
Kleis_2007, Ziambaras_2007}. Of more interest is the comparison of
calculated adsorption energies with measured heats of adsorption
\cite{Lee_2012, Poloni_2012, Chakarov_2006, Kong_2009, Zhou_2008}. As
pointed out above, IR spectroscopy can be a very powerful method to
study the loading of MOFs, but caution is neccesary. From the
theoretical side, while the full calculation of IR spectra is possible
\cite{PhysRevB.83.121402}, it is much easier and typically sufficient to
calculate the IR peak positions---this has been done in a number of
studies and shows very good agreement with experiment \cite{wikipaper50,
Kong_2009, wikipaper43}. Comparison with IR experiments has also been
made for vdW-DF calculations of small molecule diffusion
\cite{Piero_PRL}. vdW-DF calculations for an exhaustive list of elastic
and transport properties of MOFs are also compared with experiment
\cite{Kleis_2007, Langreth_2005, Londero_2010}.

\section{Examples of Successful Combined Experimental/Theoretical studies}
\label{sec:cases}

\subsection{Studying adsorption mechanisms of small molecules in MOFs}

It has been shown that MOFs with unsaturated metal centers, such as
MOF74 and HKUST-1, exhibit a fast and specific CO$_2$ absorption, which
is a desirable property for capturing applications
\cite{CO2_capt5,CO2_capt6,CO2_capt7,CO2_capt8}. Therefore, understanding
their absorption mechanisms is critical for the rational design of
improved MOFs. In this subsection we discuss and analyze the CO$_2$
absorption in MOF74. We will show that the vdW-DF approach is critical
in order to understand and correctly explain the corresponding
experimental results. As an example, we review the CO$_2$ absorption in
Zn-MOF74 and Mg-MOF74 and show how the frequency of the asymmetric and
stretching modes are modified upon absorption \cite{wikipaper43}.

\begin{figure}
\ind\includegraphics[width=2.7in]{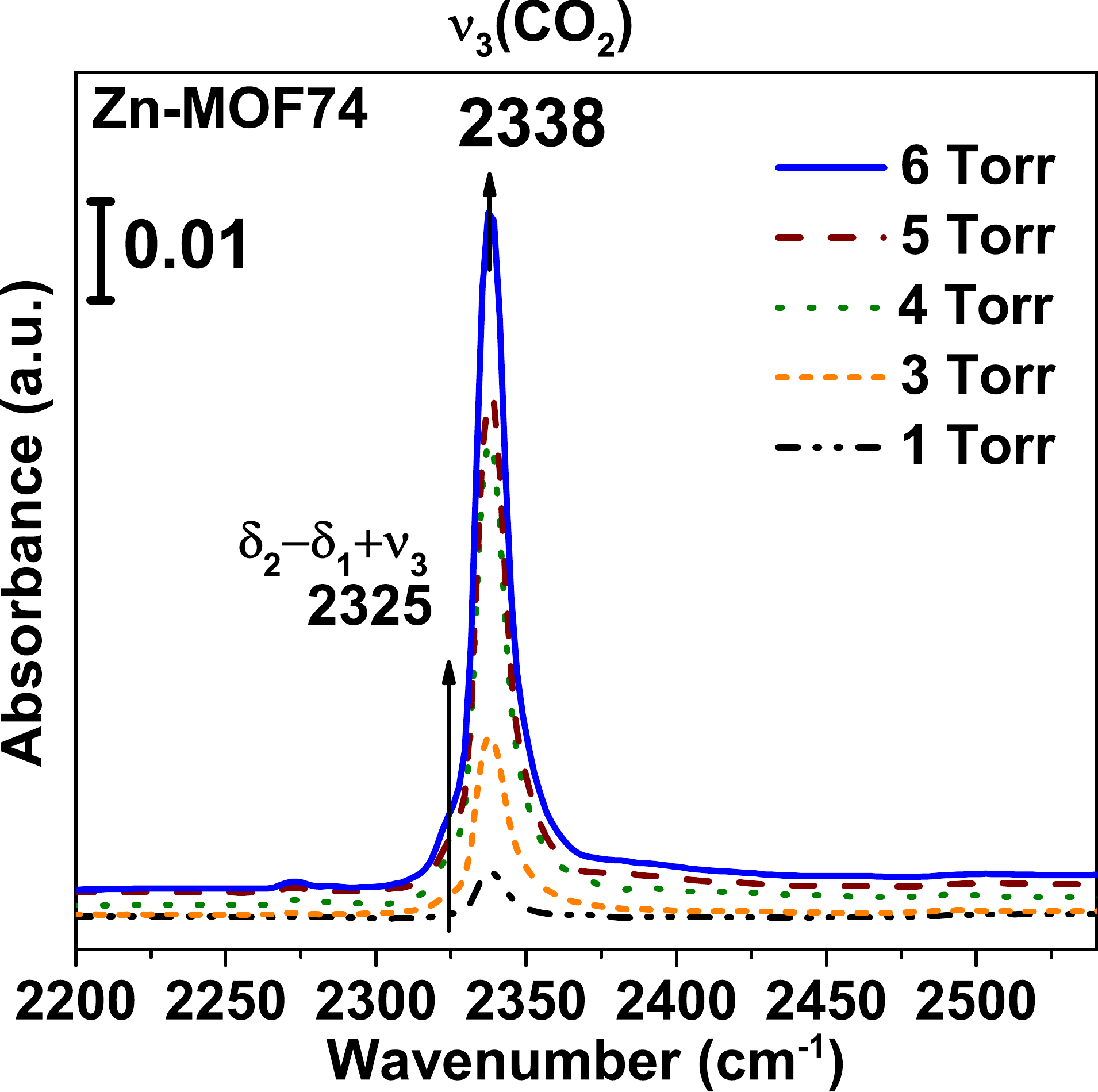}\\[3ex]
\ind\includegraphics[width=2.7in]{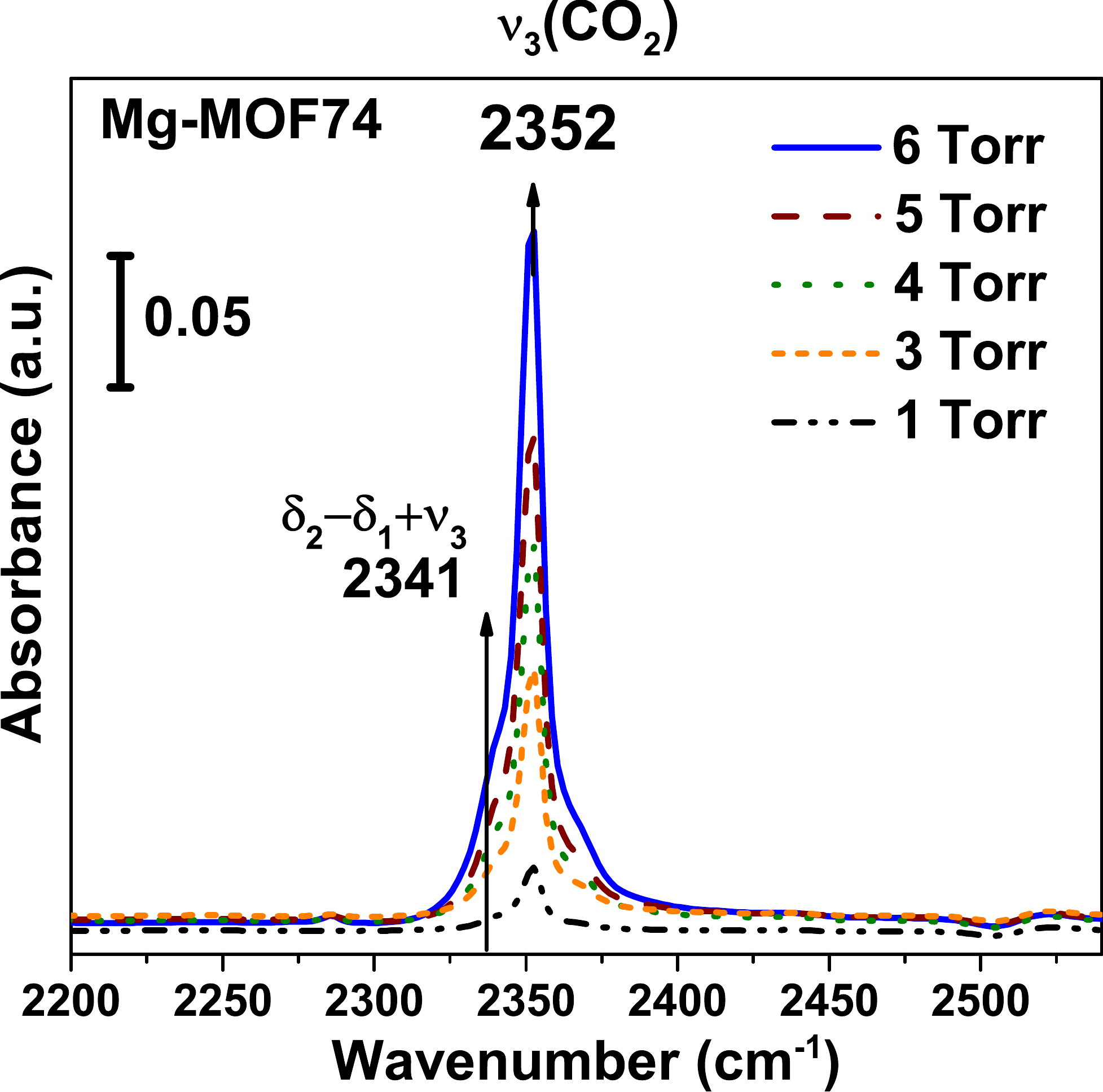}
\caption{\label{CO2_IR_shift_paper43_fig2} IR absorption spectra of
CO$_2$ absorbed into Zn-MOF74 (top) and in Mg-MOF74 (bottom) at changing
CO$_2$ pressure (1$-$6 Torr). (Reprinted with permission from Ref.
\cite{wikipaper43}. \copyright\ 2012 American Physical Society).}
\end{figure}

The experimental IR absorption spectra results in
Figure~\ref{CO2_IR_shift_paper43_fig2} show that the unperturbed
asymmetric stretch mode of CO$_2$ (2349 cm$^{-1}$) undergoes a shift of
$-11$~cm$^{-1}$ and $+3$~cm$^{-1}$ upon adsorption on Zn-MOF74 and
Mg-MOF74, respectively. But, what causes this shift? To answer this
question, \emph{ab initio} calculations were performed utilizing vdW-DF,
finding three factors contributing to this shift, i.e.\ (i) the change
in the CO$_2$ molecule length, (ii) the asymmetric distortion of the
CO$_2$ molecule, and (iii) the direct influence of the metal center.

In Table~\ref{paper43T2}, we compare the IR spectroscopy data with
results from frozen-phonon vdW-DF calculations, where the CO$_2$
molecule was adsorbed at the metal site of MOF74. In particular, the
frozen-phonon calculations for the bending mode of CO$_2$ give a change
in frequency of approximately $-9$~cm$^{-1}$ during adsorption on either
metal, in excellent agreement with the experimental results.
Furthermore, the calculations show that the asymmetric stretch mode of
the CO$_2$ molecule exhibits a red-shift of $-0.5$~cm$^{-1}$ and
$-8.1$~cm$^{-1}$ when adsorbed on Mg-MOF74 and Zn-MOF74, respectively,
in reasonable agreement with the change of $+3$~cm$^{-1}$ and
$-11$~cm$^{-1}$ measured in experiment.

\begin{table}
\caption{\label{paper43T2}Vibrational frequencies (cm$^{-1}$) of CO$_2$
physiadsorbed in MOF74. Data taken from Ref.~\cite{wikipaper43}.}
\ind\begin{tabular}{@{}llcc@{}}
\hline\hline
     & System & Bending mode ($\nu_2$) & Asym. stretch mode ($\nu_3$)  \\\hline
exp. & Free CO$_2$ & 667       & 2349 \\
     & Mg-MOF74    & 658       & 2352 \\
     & Zn-MOF74    & 658       & 2338 \\\hline
calc.& Free CO$_2$ & 646.6     & 2288.5\\
     & Mg-MOF74    & 636.6     & 2288.0 \\
     & Zn-MOF74    & 637.6     & 2280.4 \\
\hline\hline
\end{tabular}
\end{table}

According to vdW-DF calculations \cite{wikipaper43}, the CO$_2$ molecule
binds stronger to Mg-MOF74 than to Zn-MOF74, in agreement with
experimental findings. Furthermore, the distance between the metal
center and the CO$_2$ molecule is smaller in Mg-MOF74 than in Zn-MOF74.
Also, the CO$_2$ molecule experiences a larger distortion upon
adsorption in Mg-MOF74, see Table~1 in Ref.~\cite{wikipaper43}.
Therefore, it is surprising that the frequency shift of the asymmetric
stretching mode (see $\nu_3$ in Table \ref{paper43T2}) for CO$_2$ in
Mg-MOF74 is smaller compared with that in Zn-MOF74, and a deeper
investigation of what causes this peculiar result is warranted. As
mentioned above, this result can be explained with the help of theory.

We will start with the change in the molecule length: in order to
analyze this effect, phonon calculations of the free CO$_2$ molecule
were performed, where its length was set to the value when adsorbed in
the MOF, keeping the carbon atom centered. Using this approach,
frequency shifts of $-1.6$~cm$^{-1}$ and $-3.7$~cm$^{-1}$ were obtained
for the case of Mg- and Zn-MOF74, respectively. It is interesting to see
that in the case of Mg-MOF74, the molecule experiences a marginal
elongation of 0.0003~\r{A}, while in the case of Zn-MOF74 an elongation
of 0.0009~\r{A} takes place. That is, the molecule that experiences the
larger elongation, exhibits the larger red-shift, as suggested by common
sense. 

The effect corresponding to the molecule's asymmetric distortion was
studied by placing the CO$_2$ molecule exactly at the same geometry as
when adsorbed in the MOF, but removing the surrounding MOF. By doing
this, the only contributions to the change in frequency come from the
elongation of the CO$_2$ molecule and the asymmetric distortion of the
carbon atom. The former has been reported in the paragraph above, so
that the latter can easily be calculated. In this way, we find the shift
corresponding to the induced asymmetry of the CO$_2$ molecule to be
$1.1$~cm$^{-1}$ and $0.7$~cm$^{-1}$ for Mg-MOF74 and Zn-MOF74,
respectively.

Finally, the effect of the metal center was studied by placing the free,
undistorted CO$_2$ molecule at the metal adsorption site with the same
position and angle of the adsorbed system. By doing this, the change in
frequency has its highest contribution coming from the oxygen--metal
interaction. Using this configuration, the results show a frequency
shift of the asymmetric stretching mode of the CO$_2$ molecule of
$-5$~cm$^{-1}$ for the Zn-MOF74 system. On the other hand, for the
Mg-MOF74 the frequency shift has a negligible value of $-0.6$~cm$^{-1}$.
This is a striking result, since Mg and Zn have a very similar valence
structure with $3s$ and $4s$ electrons as the outermost valence states.
This result shows that the fully occupied semi-core $3d$ electrons in Zn
have an important effect on the interaction with the adsorbed CO$_2$
molecules. Similar results are found in Co- and Ni-MOF74 structures. To
shed more light on this situation, a charge-density analysis was
performed, finding a depletion of electrons around the Zn atom upon
adsorption of the CO$_2$ molecule, while this depletion was not present
for Mg-MOF74. Thus, the depletion of charge is an effect of the Zn $d$
orbitals, which, in turn, also influences the charge distribution in the
adsorbed CO$_2$ molecule. Via this mechanism, the Zn $d$ orbitals
indirectly affect the IR frequency shift of the adsorbed CO$_2$
molecule---explaining the differences between Mg-MOF74 and Zn-MOF74.

In summary, this van der Waals study of small molecule adsorption on
MOFs is driven by experimental IR data. But, it is clear that the
reasons for the observed IR frequency shifts are not necessarily
intuitive and can only be explained with the help of detailed
first-principles simulations.

\subsection{Studying the chemical stability of MOFs under humid conditions}

The stability of MOFs under humid conditions \cite{Yang_2011, Wu_2010,
Greathouse_2006, Ma_2011, Low_2009, Saha_2010} is of great importance
for the implementation of these systems in various applications and
devices. For example, the MOF5 structure is very sensitive to water and
its hydrogen uptake properties become compromised when it is exposed to
humidity in ambient air. So, how can we design new MOFs that keep their
desired properties while being water resistant? In the case of MOF5,
Yeng et al.\ \cite{Yang_2011} reported the synthesis of methyl- and
2,5-dimethyl-modified versions. By introducing methyl into the linkers,
the structure becomes less reactive to water and retains the same
hydrogen uptake properties of MOF5 up to 4 days after being exposed to
ambient air. While this is a specific case, resting on the specific
interaction of water and H$_2$ with the methyl-modified linkers, it can
easily be generalized and it is again the interaction of small
molecules---in this case water---with the MOF that is the focus of much
ongoing research.

\begin{figure}
\ind\includegraphics[width=5.15in]{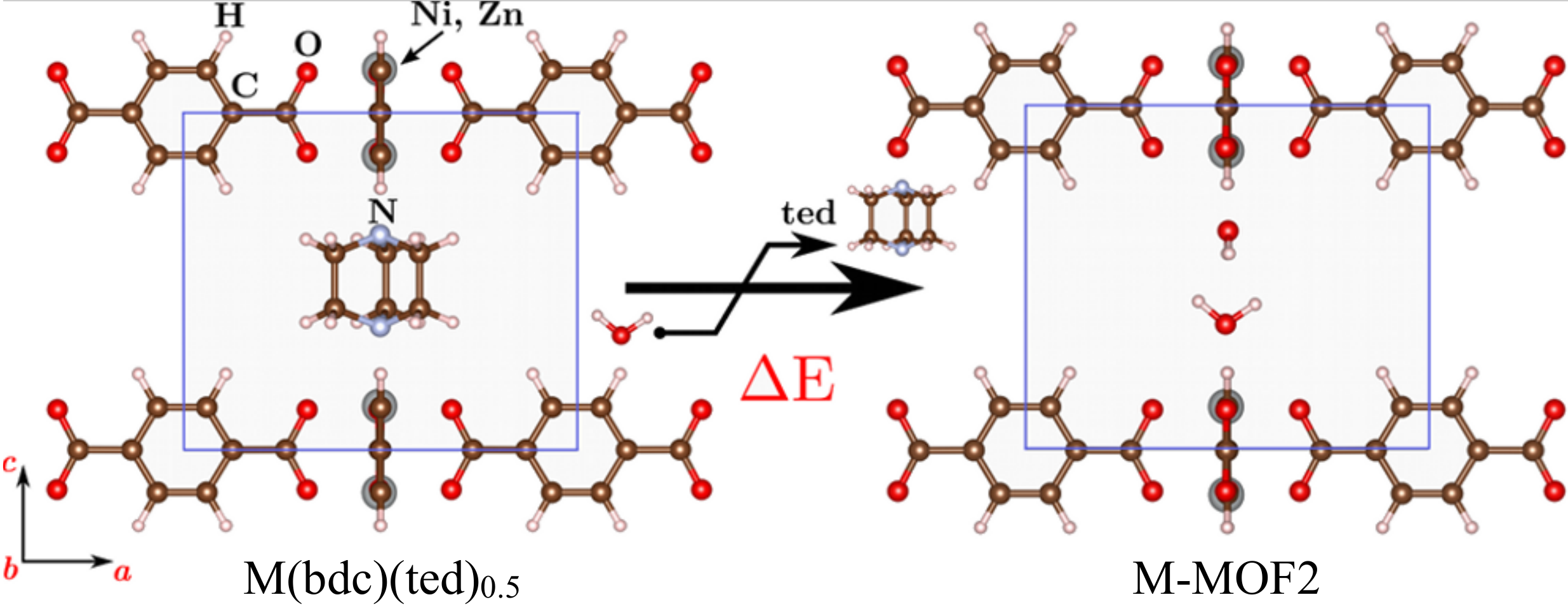}
\caption{\label{wiki_paper45_fig2} Scheme adopted for water insertion in
M(bdc)(ted)$_{0.5}$ [M = Zn, Ni], where the ted group has been
substituted by two water molecules. (Reprinted with permission from
Ref.~\cite{wikipaper45}. \copyright\ 2012 American Chemical Society).}
\end{figure}

In this subsection, we review efforts to understand the MOF--water
interaction, using as an example the prototypical metal organic
framework M(bdc)(ted)$_{0.5}$ [M = Cu, Zn, Ni, Co; bdc =
1,4-benzenedicarboxylate; ted = triethylenediamine]. This MOF has shown
promising properties towards the adsorption of gases, such as H$_2$,
CO$_2$, and CH$_4$ \cite{Chen_2010, Lee_2007, Dybtsev_2004}.
M(bdc)(ted)$_{0.5}$ exhibits thermal stability up to 282$^{\circ}$C, is
highly porous, the H$_2$ adsorption is exceptionally high, and it can
also adsorb a large amount of hydrocarbons. This system was first
synthesized and reported by Dybtsev et al.\ in Ref.~\cite{Dybtsev_2004}
and we will review here its water stability, as originally studied in
Ref.~\cite{wikipaper45}. A characteristic building block of this
particular MOF is the incorporated ``paddle wheel'' building-block ted
(triethylenediamine), which acts as a linker. In the presence of water,
this paddle wheel structure can be extracted from the framework and be
replaced by water molecules, forming M-MOF2 (we will refer to it as
MOF2), as can be seen in Figure~\ref{wiki_paper45_fig2}. Obviously, with
its normal linker missing, the M(bdc)(ted)$_{0.5}$ structure looses
stability and, in most cases, undergoes an non-reversible phase
transition. 

\begin{figure}
\ind\includegraphics[width=5.15in]{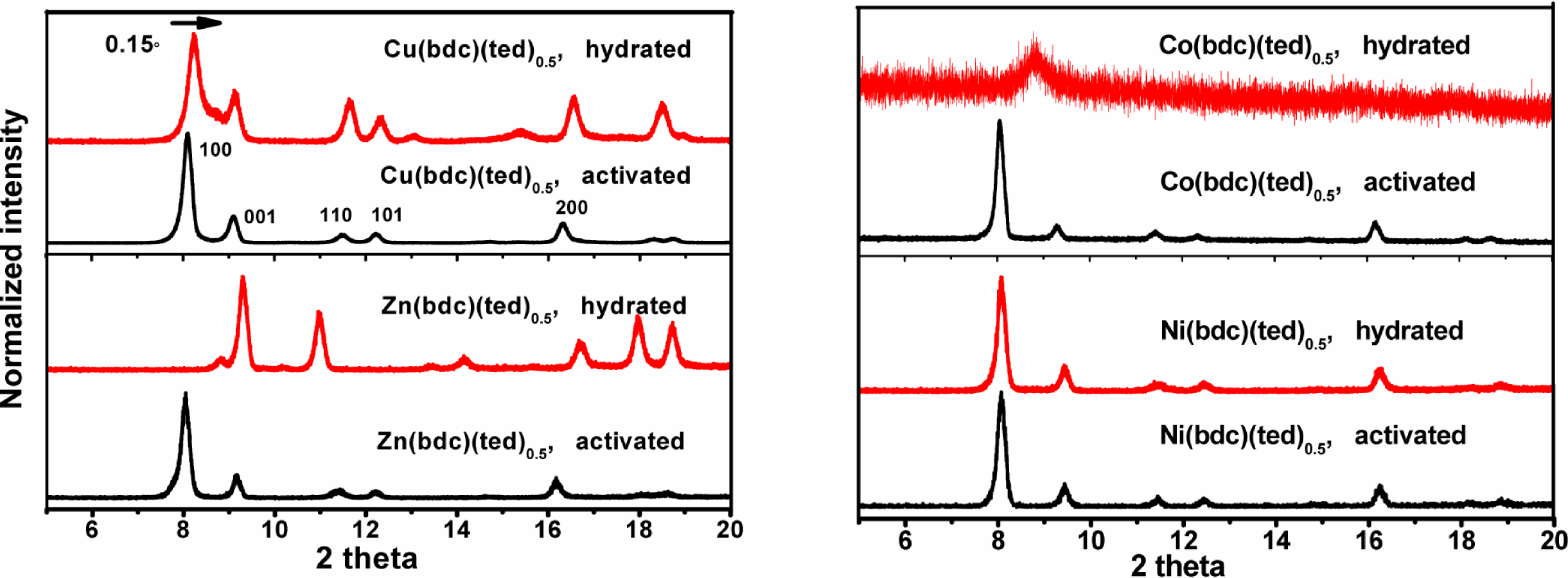}
\caption{\label{Tan_2012_fig8}Powder X-ray patterns of activated
(pristine) and hydrated M(bdc)(ted)$_{0.5}$ [M = Cu, Zn, Co and Ni]
after exposure to 9.5 Torr of D$_2$O vapor. (Reprinted with permission
from Ref.~\cite{wikipaper45}. \copyright\ 2012 American Chemical
Society).}
\end{figure}

Figure~\ref{Tan_2012_fig8} shows the powder X-ray diffraction (XRD)
pattern of four hydrated M(bdc)(ted)$_{0.5}$ systems [M = Cu, Zn, Co,
and Ni] after exposure to 9.5 Torr of D$_2$O vapor and the corresponding
activated (pristine) M(bdc)(ted)$_{0.5}$ samples. Concerning the
Cu(bdc)(ted)$_{0.5}$ system, the XRD pattern confirms that the system is
stable after exposure to D$_2$O gas up to a pressure of 6 Torr, see
Figure~S10 in the supporting information of Ref.~\cite{wikipaper45}.
However, the top left panel of Figure~\ref{Tan_2012_fig8} shows that all
the X-ray peaks are shifted to higher angles, except for the [001] peak,
indicating that the Cu(bdc)(ted)$_{0.5}$ system is partially hydrolyzed
by the D$_2$O molecules. Even though the structure is only partially
hydrolyzed, the original Cu(bdc)(ted)$_{0.5}$ structure cannot be
recovered after evacuation of water at a temperature of 150$^\circ$C. In
contrast, the left bottom panel of Figure~\ref{Tan_2012_fig8} clearly
indicates that the Zn(bdc)(ted)$_{0.5}$ system transforms into MOF2
after hydration. This transformation starts with the detachment of the
ted group and the subsequent bonding of the D$_2$O molecules to the
Zn$^{2+}$ apical sites of the paddle-wheel building units through their
oxygen atoms. Concerning the Ni(bdc)(ted)$_{0.5}$ and
Co(bdc)(ted)$_{0.5}$ systems under humid conditions, the right bottom
and right upper panels of Figure~\ref{Tan_2012_fig8} indicate that the
Ni(bdc)(ted)$_{0.5}$ maintains its structure after been exposed to 9.5
Torr of D$_2$O vapor, while the Co(bdc)(ted)$_{0.5}$ is completely
destroyed after exposure. Furthermore, the Co(bdc)(ted)$_{0.5}$
structure cannot be recovered after annealing in vacuum up to
150$^\circ$C, see Figure~S13 in the supplemental material of
Ref.~\cite{wikipaper45}.

In order to explain the previous experimental results and give a clear
explanation of how water interacts with the M(bdc)(ted)$_{0.5}$, we
review computational results obtained in Ref.~\cite{wikipaper45}
concerning the Ni(bdc)(ted)$_{0.5}$ and Zn(bdc)(ted)$_{0.5}$ systems.
The energy $\Delta E$ needed to extract the paddle wheel and replace it
with water molecules was calculated using the vdW-DF formalism as
\begin{eqnarray}
\label{wikipaper45_Eq_1}
\Delta E_{\mathrm{M(bdc)(ted)}_{0.5}} &=& E[\mathrm{M(bdc)(ted)}_{0.5}]+\,E[n\,\mathrm{H}_{2}\mathrm{O}] \nonumber \\
&-& E[\mathrm{MOF2}+n\,\mathrm{H}_{2}\mathrm{O}] -1/2\,E[(\mathrm{ted})]\;,
\end{eqnarray}
where $n$ is the number of water molecules in the MOF,
$E[\mathrm{M(bdc)(ted)}_{0.5}]$ is the energy of the MOF with no water
molecules in it (as seen in the left panel of
Figure~\ref{wiki_paper45_fig2}), $E[n\,\mathrm{H}_{2}\mathrm{O}]$ is the
energy of $n$ water molecules,
$E[\mathrm{MOF2}+n\,\mathrm{H}_{2}\mathrm{O}]$ is the energy of the
M(bdc)(ted)$_{0.5}$ where the ``exiting" ted has been replaced with $n$
water molecules (right panel of Figure~\ref{wiki_paper45_fig2}) and
$E[(\mathrm{ted})]$ is the energy of the ted. Table~\ref{wiki_paper45T2}
shows the energies required to substitute the ted in the Zn and
Ni(bdc)(ted)$_{0.5}$ structures by 2, 4, 6, 8, and 10 water molecules.
Note that negative $\Delta E$ values indicate that the replacement is
energetically favorable. The table shows that Ni(bdc)(ted)$_{0.5}$ is
more resistant to water than Zn(bdc)(ted)$_{0.5}$, as found in the
spectra in Figure~\ref{Tan_2012_fig8}, and the hydration of the latter
is a spontaneous process. This is due to the strong H bonds between the
water molecules, which stabilizes the coordination of the Zn metal
centers. On the other hand, in the case of Ni(bdc)(ted)$_{0.5}$, 
$\Delta E_{\mathrm{M(bdc)(ted)}_{0.5}}$ becomes negative only 
when the number of water molecules is 6 or
greater. 

Alternatively, one can calculate the energy 
$\Delta E_{\mbox{\scriptsize M-MOF2}}$ required for hydration of the 
MOF2 structure with $n$ water molecules, using:
\begin{eqnarray}
\label{wikipaper45_Eq_2}
\Delta E_{\mbox{\scriptsize M-MOF2}} &=& E[\mathrm{MOF2}]+\,E[n\,\mathrm{H}_{2}\mathrm{O}] \nonumber \\
&-& E[\mathrm{MOF2}+n\,\mathrm{H}_{2}\mathrm{O}]\;.
\end{eqnarray}
Here, $E[\mathrm{MOF2}]$ is the energy of the M(bdc)(ted)$_{0.5}$, where
the ted has been replaced by two molecules of water; the other terms in
the equation have been previously defined, see
Equation~\ref{wikipaper45_Eq_1}. The right hand side of
Table~\ref{wiki_paper45T2} shows that for MOF2 the hydration of the Zn
and Ni systems is a spontaneous process with an energy gain of
approximately $-55$~kJ mol$^{-1}$ cell$^{-1}$ for higher loadings. This
trend is almost independent of the metal.

In conclusion, the computational results explain the experimental
findings in Ref.~\cite{wikipaper45}, indicating that the structural
stability of the system depends on the amount of water present in the
MOF. At lower loadings the system is stable, while at higher loadings
the interaction of water with the paddle wheel leads to the irreversible
decomposition of the structure.

\begin{table}
\caption{\label{wiki_paper45T2}Computed $\Delta E_{\mathrm{M(bdc)(ted)}_{0.5}}$ 
and $\Delta E_{\mbox{\scriptsize M-MOF2}}$ (kJ mol$^{-1}$ cell$^{-1}$) 
as a function of the number of water
molecules per cell. Note that the basic MOF2 structure already contains
two water molecules. Data taken from Ref. \cite{wikipaper45}.}
\ind
\begin{tabular}{@{}lrrrrrrr@{}}
\hline\hline
& \multicolumn{3}{c}{$\Delta E_{\mathrm{M(bdc)(ted)}_{0.5}}$}& &\multicolumn{3}{c}{$\Delta E_{\mbox{\scriptsize M-MOF2}}$ }\\[0.5ex]
\cline{2-4}\cline{6-8}
H$_2$O/cell  &Zn &Ni &$ \Delta $ &  &Zn &Ni &$ \Delta $ \\
\hline
2  & 43.1    & 85.5    &42.4 &  &---    &---     &---  \\
4  &--5.3    &4.2      &9.5  &  &--53.6 &--77.1  &--23.5 \\
6  &--21.4   &--17.1   &4.3  &  &--53.7 &--68.4  &--14.7 \\
8  &--31.3   &--24.0   &7.3  &  &--56.1 &--52.4  &3.7  \\
10 &--35.0   &--45.2   &--10.2 &  &--54.5 &--55.3&--0.8  \\
\hline\hline
\end{tabular}
\end{table}

\subsection{Studying the formation of water clusters in fluorinated MOFs}

The large internal surface area of MOFs makes them ideal for catalysis
and fuel cell applications, which have attracted a surge of interest
\cite{Kitagawa_FC, Shultz_MOFcatal, Farha_MOFcatal2, Lee_MOFcatal3,
mof_app, Xamena_MOFphotoCatal}. While some progress has been made---for
example, Hurd et al.\ \cite{PEM_MOF} show intriguing results for
$\beta$-PCMOF2 (proton conducting metal organic framework 2), capable of
proton transport under anhydrous conditions at
$150^{\circ}\mathrm{C}$---in general, the low hydrothermal and chemical
stability of MOFs prevents their implementation in catalytic and
fuel-cells systems. In the recent past, thus, concerted efforts have
focused on increasing the hydrothermal and chemical stability of
MOFs \cite{Ma_2008, sun_2005, Low_2009}.

A promising approach to increase the chemical and hydrothermal stability
is fluorinated MOFs (FMOFs), where the H atoms have been replaced by F
atoms \cite{Yang_FMOF, Yang_FMOF2, Yan_2011}. Yang et al.\ report
interesting results for FMOF1, showing that the hydrogen-desorption
isotherm does not follow the path of the adsorption isotherm
\cite{Yang_FMOF}, in fact, it shows an abrupt drop in the adsorption
density at 14 bar. The authors highlight the fact that this behavior
would allow FMOF1 to adsorb H$_2$ at high pressures and stored it at low
pressures.

In general, the walls of FMOF systems are hydrophobic, leading to an
interesting side effect: the weak interaction of water molecules with
the FMOF enhances the creation of water clusters inside its pores. In
this subsection, we review the formation and behavior of water clusters
inside FMOF1, as reported in Ref.~\cite{wikipaper55}. As in previous
sections, an understanding of the weak molecular interactions inside
this system was gained by a combination of vdW-DF calculations and IR
absorption spectra of water exposed FMOF1 as a function of pressure.
Note that the interaction between water molecules has a significant van
der Waals component, which is well captured with vdW-DF \cite{Kolb2011},
while the electrostatic interaction is suppressed by the wall
hydrophobicity of FMOF1.

\begin{figure}
\ind\includegraphics[width=3.5in]{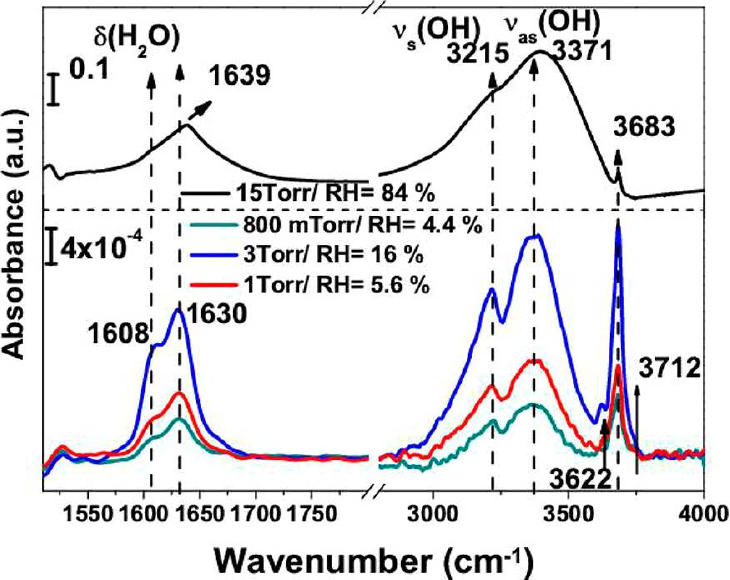}
\caption{\label{Figure_2_wikipaper55} IR absorption spectra of water
exposure in FMOF1 as a function of pressure. Absorption spectra are
referenced to dehydrated FMOF1 in vacuum. The top panel shows exposure
at 15~Torr, while the bottom part shows exposure at lower pressures
(800~mTorr to 3~Torr). (Reprinted with permission from Ref.
\cite{wikipaper55}. \copyright\ 2013 American Chemical Society).} 
\end{figure}

\begin{figure}
\ind\includegraphics[width=3.5in]{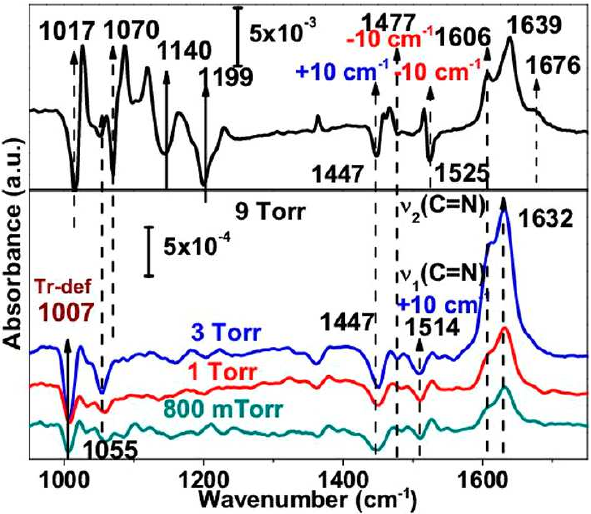}
\caption{\label{Figure_3_wikipaper55}IR absorption spectra of H$_2$O
adsorbed in FMOF1 showing the bending modes of adsorbed water as a
function of pressure. Top part shows IR absorption spectrum at 9 Torr.
(Reprinted with permission from Ref. \cite{wikipaper55}. \copyright\
2013 American Chemical Society).}
\end{figure}

Experimental isotherm measurements of FMOF1 show that the adsorption of
water is negligible compared to the water adsorption in other systems
\cite{Yan_2011}. Furthermore, at low water pressures (800 mTorr to 3
Torr), the IR adsorption measurements of H$_2$O adsorbed on FMOF1 show
two peaks corresponding to red ($-13$~cm$^{-1}$) and blue-shifts
($+9$~cm$^{-1}$) of the unperturbed scissor vibration mode
(1621~cm$^{-1}$) of the water molecule, as can be seen
Figure~\ref{Figure_2_wikipaper55}. On the other hand, as the pressure is
increased to 9 Torr, new peaks associated with scissor vibration modes
appear at 1639~cm$^{-1}$ and 1676~cm$^{-1}$, as can be seen in the top
panel of Figure~\ref{Figure_3_wikipaper55}.

\begin{figure}
\ind\includegraphics[width=4in]{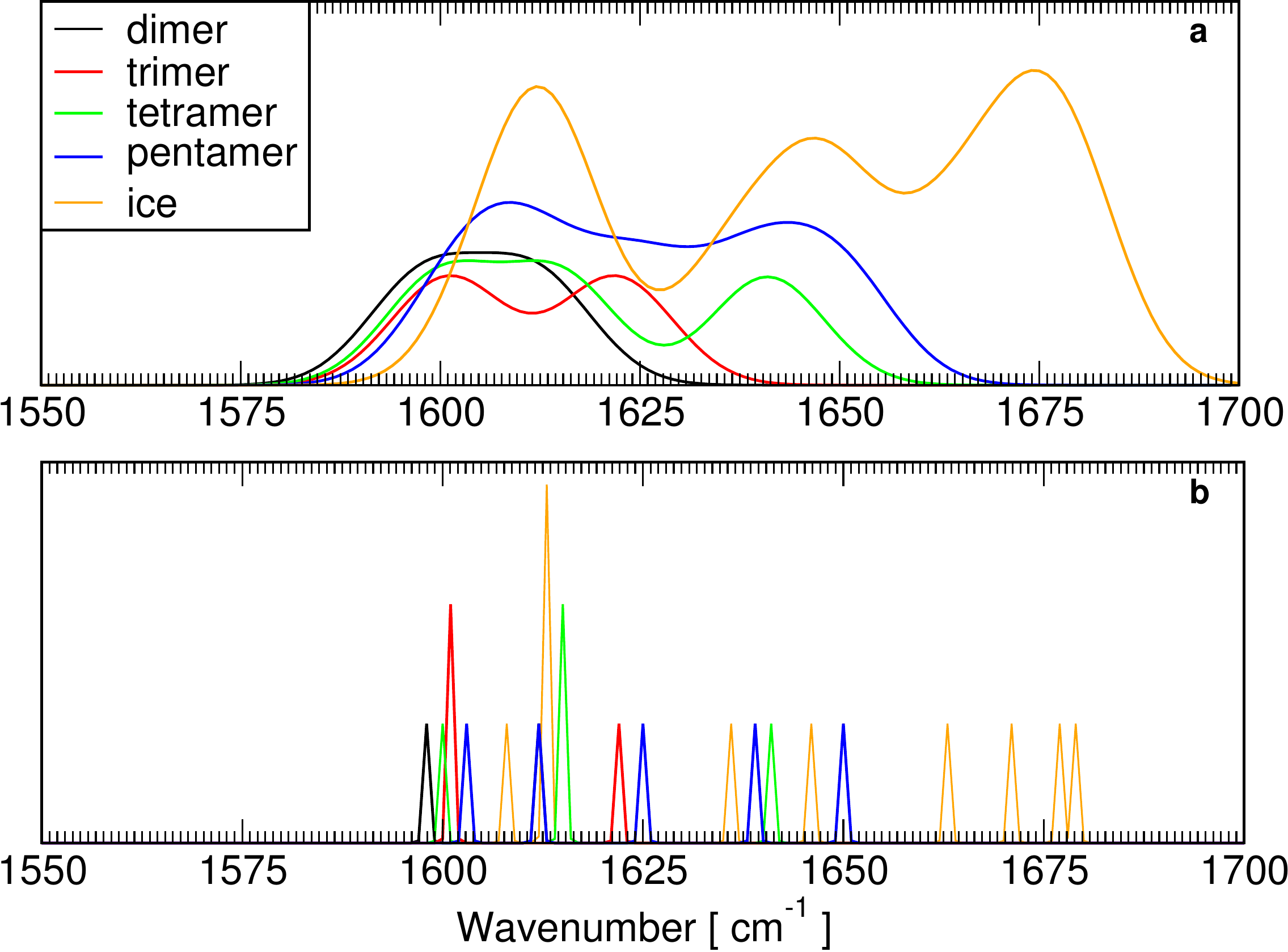}
\caption{\label{Figure_6_wikipaper55}a) Gaussian convolution (with
bandwidth of 20~cm$^{-1}$) of bending mode frequencies for various
cluster sizes. b) Single frequency values represented by peaks of
1~cm$^{-1}$ width, as reported by previous vdW-DF calculations on
gas-phase water clusters \cite{Kolb2011}. (Reprinted with permission
from Ref. \cite{wikipaper55}. \copyright\ 2013 American Chemical
Society).}
\end{figure}

In order to elucidate the appearance and nature of these peaks, vdW-DF
vibration calculations were performed for various water clusters, i.e.\
the water dimer, trimer, tetramer, pentamer, and ice; the results are
shown in Figure~\ref{Figure_6_wikipaper55}.
Figure~\ref{Figure_6_wikipaper55}a) shows the calculated modes
convoluted by Gaussian functions of 20~cm$^{-1}$ bandwidth, while panel
b) shows single frequency values represented by peaks of 1~cm$^{-1}$
width. As expected, from the figure it can be seen that the bigger the
water cluster, the higher the number of scissors modes. It is also
important to note that for pressures under 3~Torr, the scissor
vibrational modes in Figure~\ref{Figure_2_wikipaper55} span from
1600~cm$^{-1}$ to 1650~cm$^{-1}$. This matches the theoretical frequency
windows of both the tetramer and pentamer, as seen in the top panel of
Figure~\ref{Figure_6_wikipaper55}. It follows that the water clusters
formed inside FMOF1 under low pressures ($<3$~Torr) are comprised of no
more than 5 water molecules. This conclusion is also supported by the
water adsorption energies on the FMOF1, see Table~2 in
Ref.~\cite{wikipaper55}. Note that, in principle, up to 61 water
molecules can be accommodated inside the pores of FMOF1. On the other
hand, the experimentally observed peak located at 1676~cm$^{-1}$ in
Figure~\ref{Figure_3_wikipaper55} can be associate with hydrogen-bonded
water molecules or water clusters larger than five water molecules---see
the orange line in the top panel of Figure~\ref{Figure_6_wikipaper55}.
It is important to note that this peak is only visible at high
pressures.

In summary, while the IR spectroscopy data of water-exposed FMOF1 showed
the appearance of new peaks, it was only with the help of vdW-DF
calculations that a clear assignment to particular water clusters could
be made. Note that this finding is likely to have a tremendous impact on
atmospheric sciences, which seek to study the existence and properties
of such clusters. In the normal atmosphere, water cluster concentration
decays exponentially with the aggregate size, making clusters larger
than the trimer often difficult to observe. FMOF1 solves this problem
and provides a simple environment to create and confine even larger
clusters.

\subsection{Studying small molecule diffusion in MOFs}

MOFs have attracted a lot of attention due their promising properties
concerning the storage of hydrogen and capture of
CO$_2$ \cite{Pera_2013}, among others. However, for the effectiveness of
all such applications, it is necessary to get guest molecules deep into
the bulk of the MOF, or vice versa, have them diffuse out. As such, the
diffusivity of the guest molecule through the porous material plays a
major role in these processes and is critical for the understanding and
rational design of new MOFs. The topic of small molecule diffusion in
MOFs has thus been the target of many theoretical studies
\cite{Skoulidas_2005, Haldoupis_2012, Yang_2005, Amirjalayer_2007,
Skoulidas_2004, Haldoupis_2010, Liu_2008}. For example, in
Ref.~\cite{Haldoupis_2010} Haldoupis et al.\ identified key elements in
the MOF's pore structure and via molecular dynamic simulations they were
able to predict the Henry constant and the activation energy for several
guest molecules. In particular, the authors were able to identify
several materials with promising properties towards the separation of
gases, such as H$_2$, CO$_2$, and CH$_4$. However, in their study, the
authors assume that the MOFs are rigid structures, which can be a
serious limitation, as we know that some MOFs experience a significant
change in their structure upon adsorption of the guess molecules or
other external stimuli due to their high flexibility.

In this subsection we review a combined {\it in situ} IR/vdW-DF study of small
molecule diffusion in Mg-MOF74, as described in \cite{wikipaper51}.
MOF74 was chosen for this study due to its unsaturated metal centers,
which makes it highly reactive towards the adsorption of small
molecules. Furthermore, Mg-MOF74 has shown promising properties towards
the adsorption of CO$_2$ compared to other MOFs.

We start by showing results concerning the adsorption energies of H$_2$,
CO$_2$, and H$_2$O in the Mg-MOF74 structure, see
Table~\ref{PRL_110_026102_T1}. This table shows that for low to moderate
loadings the interaction between adsorbate molecules is negligible,
except for H$_2$O adsorption, where the repulsion between the H atoms of
the water molecules slightly debilitates the H$_2$O binding to the MOF.
The adsorption energies for the adsorption of H$_2$ and CO$_2$, obtained
using the vdW-DF approach, are in excellent agreement with the
experimental values of $-0.11$~$\pm$ 0.003~eV \cite{Zhou_2008} and
$-0.49$~$\pm$ 0.010~eV \cite{CO2_capt7}, respectively. Although not the
focus of that particular study, Table~\ref{PRL_110_026102_T1} also
reveals a common problem of many MOFs: the adsorption energy of water
(due to its large dipole moment) is typically significantly higher
compared to e.g.\ H$_2$ and CO$_2$. Thus, the presence of even small
traces of water is a serious impediment upon possible applications and
devices, as anticipated in the previous section. Details about this
problem are discussed in Ref.~\cite{Canepa2013}. In addition to
adsorption energies, calculations of the vibrational spectra show a
frequency change after adsorption of $\Delta \nu_{\rm H_2} =
-30$~cm$^{-1}$, $\Delta \nu_{\rm CO_2} = -13$~cm$^{-1}$, and $\Delta
\nu_{\rm H_2O} = -103$~cm$^{-1}$, in remarkable agreement with the IR
spectroscopy measurement of $\Delta \nu_{\rm H_2} = -36$~cm$^{-1}$,
$\Delta \nu_{\rm CO_2} = -8$~cm$^{-1}$, and $\Delta \nu_{\rm H_2O} =
-99$~cm$^{-1}$. Experimentally, there is also a small difference in
frequency change between low and high loading, resulting in a red-shift
of $-3$~cm$^{-1}$ and $-15$~cm$^{-1}$ for the asymmetric stretch modes
of CO$_2$ and H$_2$O (see Supplemental Material of
Ref.~\cite{wikipaper51}). Computationally, we find $\Delta\nu_{\rm CO_2}
= -1$~cm$^{-1}$ and $\Delta\nu_{\rm H_2O} = -18$~cm$^{-1}$, again in
excellent agreement with experiment.

\begin{table}
\caption{\label{PRL_110_026102_T1}Adsorption energies $\Delta E$ of
molecules in Mg-MOF74 in eV. Two different loading are considered, i.e.\
one molecule per unit cell (low loading) and 6 molecules per unit cell
(high loading). In addition, adsorption energies corrected for the
zero-point energy ($\Delta E_{\rm ZPE}$) and thermal contribution
($\Delta H_{298}$ at 298 K) are given in eV. Data taken from Ref.
\cite{wikipaper51}.}
\ind
\begin{tabular}{@{}l c c c r@{}}
\hline\hline
Molecule   & Loading & $\Delta E$ & $\Delta E_{\rm ZPE}$ & $H_{298}$\\ 
\hline
H$_2$      & 1       &--0.15       & --0.15            & --0.15  \\
           & 6       &--0.16       & --0.16            & --0.16  \\
CO$_2$     & 1       &--0.50       & --0.49            & --0.50  \\
           & 6       &--0.50       & --0.49            & --0.50  \\
H$_2$O     & 1       &--0.79       & --0.76            & --0.76  \\
           & 6       &--0.76       & --0.73            & --0.73  \\
\hline\hline
\end{tabular}
\end{table}

\begin{figure}
\ind\includegraphics[width=2.25in]{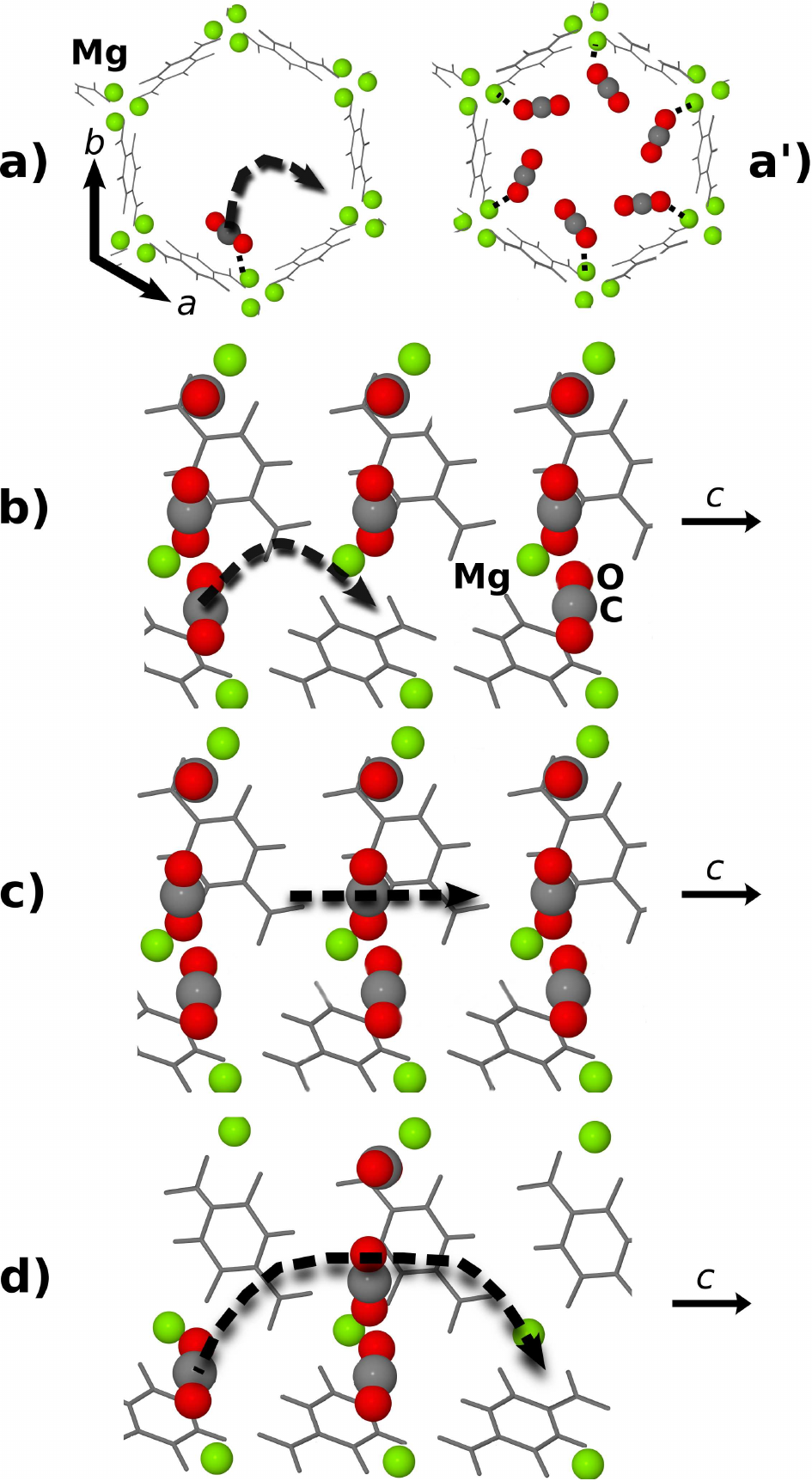}
\caption{\label{Fig1_PRL_110_026102}Graphical representation of the
diffusion mechanisms considered in this study, shown for the case of
CO$_2$. a) and a') are views directly along the $c$-axis of the
hexagonal Mg-MOF74 cell, where one (low loading) and six CO$_2$ (high
loading) are adsorbed. b), c), and d) are views perpendicular to the
$c$-axis. In panel b) the guest molecule, adsorbed on a metal center,
diffuses along the $c$-axis to the next metal center. In panel c) the guest
molecule travels along the center of the MOF channel, while all the
metal centers are occupied by the same type of guest molecule. In panel d)
the guest molecule, adsorbed on one of the metal centers, travels along
the $c$-axis through a barrier made of other adsorbed molecules and is
adsorbed again at the equivalent metal center two unit cells further down.
Dashed lines indicate the diffusion paths. (Reprinted with
permission from Ref. \cite{wikipaper51}. \copyright\ 2012 American
Physical Society).}
\end{figure}

The diffusion of small molecules (H$_2$, CO$_2$, and H$_2$O) through the
MOF is a complex process. An appropriate description of such processes
typically requires computationally expensive first-principle molecular
dynamic simulations. However, here we were able to avoid the use of
molecular dynamics by finding four different diffusion paths that
capture important molecular transport mechanisms responsible for the
macroscopic diffusion of H$_2$, CO$_2$, and H$_2$O in the MOF structure.
These four paths are: a) The guest molecule, adsorbed on one metal
center, travels circularly from one metal center to the next. Note that
this mechanism is not responsible for molecular transport into the MOF,
but nevertheless is an important process for redistributing the
molecular load. b) The guest molecule, adsorbed on the metal center,
diffuses along the $c$-axis to the next metal center. c) The guest
molecule travels through the center of the MOFs channel, where all the
metal centers are occupied by the same type of guest molecules. And, d)
the guest molecule, adsorbed on one of the metal centers, travels along
the $c$-axis through a barrier made by adsorbed molecules and is
adsorbed at the equivalent metal center two unit cell further down. See
Figure~\ref{Fig1_PRL_110_026102} for a graphical representation of these
four diffusion paths. For these paths, diffusion barriers were then
calculated utilizing vdW-DF combined with the climbing-image nudged
elastic band (NEB) approach.

\begin{figure}
\ind\includegraphics[width=3.5in]{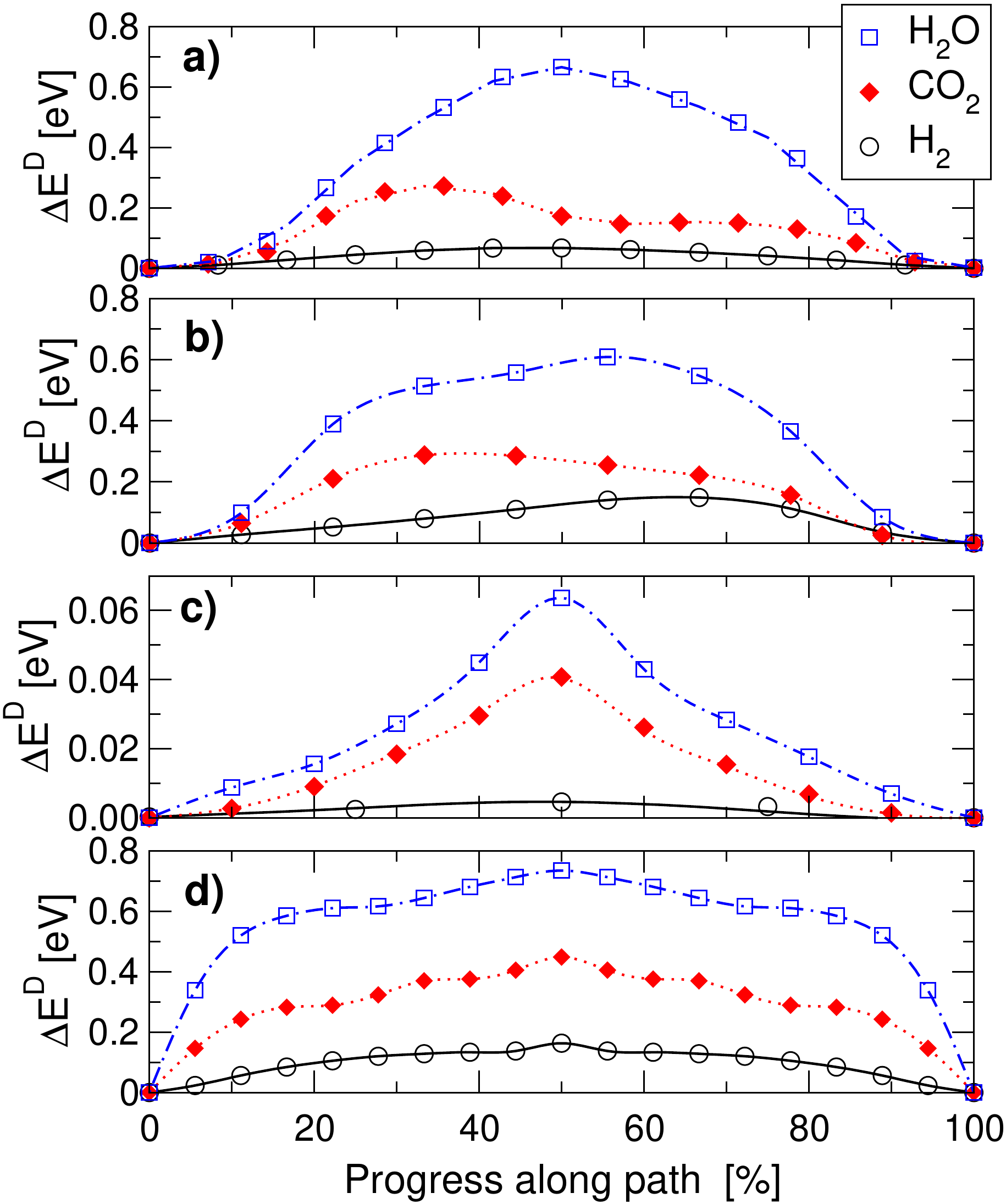}
\caption{\label{Fig2_PRL_110_026102}Diffusion profiles (in eV) for the
diffusion processes of H$_2$, CO$_2$, and H$_2$O in Mg-MOF74 according
to the mechanisms in Figure~\ref{Fig1_PRL_110_026102}. (Reprinted with
permission from Ref. \cite{wikipaper51}. \copyright\ 2012 American
Physical Society).}
\end{figure}

The energy barriers of the four diffusion paths are plotted in
Figure~\ref{Fig2_PRL_110_026102}. Note that diffusion barriers corrected
for the zero-point energy were also calculated, but are not reproduced
here. From the figure it can be seen that water has the highest energy
barrier for diffusion. Again, the presence of water inside the MOF is a
serious issue, as the barrier for it to diffuse out is rather large. As
expected, the energy barriers are comparable to the adsorption energies.
In panel a), it can be seen that a local minimum is located at 58$\%$ of
the path for CO$_2$ diffusion. This local minimum has its origins in the
presence of a secondary adsorption site in the MOF. Due to its low depth
(5~meV), the secondary adsorption site is only occupied at high
loadings. This secondary adsorption site for the CO$_2$ molecule was
first reported by Queen et al.\ in Ref.~\cite{Queen_2011}, where the
authors conducted neutron powder diffraction experiments on the Mg-MOF74
as a function of the CO$_2$ loading. Paths b), c), and d) aim to
simulate the diffusion of the guest molecule into the MOF. Note that the
diffusion barrier in path c) are ten times lower than the ones obtained
in paths a), b) and d). This indicates that the interaction between the
guest molecules in the middle of the channel and the ones adsorbed at
the metals sites is small. Furthermore, it is important to highlight
that the diffusion energy barrier of CO$_2$ in
Figure~\ref{Fig2_PRL_110_026102}c), i.e.\
0.04~eV, becomes 0.03~eV when corrected for the zero-point energy. This
value is in excellent agreement with the 0.03~eV energy barrier measured
experimentally by Bao et al.\ in Ref.~\cite{Bao_2011}.

In addition to vdW-DF calculations, {\it in situ} IR time-resolved
spectroscopy measurements of the diffusion of CO$_2$ and H$_2$O in
Mg-MOF74 were performed. When the experiments were first performed, the
results were difficult to understand. In the case of CO$_2$, at first we
observed a red-shift in the vibration frequency (asymmetric mode) of the
guest molecule.  As time passes, the IR spectrum measurements show a
second shift (blue-shift) in the vibration frequencies of the guest
molecules, leading to the original IR spectrum. The analogous behavior
is observed for the corresponding experiment with H$_2$O. With the help
of theory, we were able to construct a model that explains these
effects. At first, molecules entering the MOF mostly adsorb in the pores
close to the surface and ``clog'' those. This causes the first
experimentally observed red-shift. Those pores become highly loaded,
which we were able to deduce from the calculated difference in frequency
shift from the low- and high-loading situations. Then, after some time,
molecules start to diffuse deeper into the MOF using mechanism c),
diffusing from pores with high concentration of guest molecules to pores
with lower concentration. This results in the second shift observed,
i.e.\ blue shifting back to the original spectrum.

To test our model, we compare the experimental timescale for the CO$_2$
and H$_2$O cases. The experiments show that it takes approximately two
hours for the H$_2$O molecules and 22 minutes for the CO$_2$ molecules
to go from the high-loading regime to the low-loading regime. The ratio
between these two times is 5.45. On the other hand, having calculated
the corresponding diffusion barriers (and calculating the
pre-exponential factor in the Arrhenius equation with the help of
transition-state theory), we can compute the same ratio and find purely
based on our \emph{ab initio} calculations a value of 5.43, validating
our theoretical accuracy and transport model.

In summary, as in the previous subsections, only the combination of
experiment and theory was able to present a complete picture of small
molecule diffusion in MOF74. The theoretical atomistic model for the
molecular transport explains experimental IR macroscopic evidence. The
simulations also clarify the two-state mechanism, observed
experimentally, which controls the macroscopic diffusion of these
molecules in MOF74.

\section{Summary and Outlook}\label{sec:summary}

In this work, we have shown several examples of how the synergy of IR
and Raman spectroscopy techniques, together with \emph{ab initio}
calculations at the DFT level utilizing vdW-DF, allow us to give a
complete description of the van der Waals binding and interaction
between guest molecules and the MOF. While originally many studies of
MOFs focused on adsorption of H$_2$ and CO$_2$, at the moment we see a
vast expansion of this field, including many other molecules of
interest, such as SO$_2$ and NO$_2$ \cite{Tan_2013, Ebrahim_2013,
Yu_2012, Yu_2013, Ding_2012}. Interesting effects are also being
studied, such as a pressure dependent gate opening of MOFs
\cite{wikipaper47, Coudert_2009, Canan_2010, Pera_2012} and the response
of MOFs to a variety of external stimuli. Due to the versatile
building-block nature of MOFs, an almost innumerable number of MOFs
might exist. But, more fundamental research is necessary to understand
their properties and tailor them according to our needs. Nonetheless,
and at this point we probably only have seen a glimpse of their
applicability for future applications and devices.

\section*{Acknowledgment}
This work was entirely supported by Department of Energy Grant No.\
DE-FG02-08ER46491.

\section*{References}
\bibliography{biblio}
\bibliographystyle{unsrt}

\end{document}